\documentstyle[12pt]{article}
\begin{document}

\author{C. S. Unnikrishnan$^{1,2,3,a}$, G. T. Gillies$^{4,b}$, and R. C. Ritter$%
^{5,c}$ \\
%EndAName
$^{1}${\it Gravitation Group, Tata Institute of Fundamental }Research, \\
{\it Homi Bhabha Road, Mumbai - 400 005, India}\\
$^{2}${\it Laboratoire Kastler Brossel, Ecole Normale Sup\'{e}rieure, }\\
{\it \ 24 rue Lhomond, 75231 Paris Cedex 05, France}\\
{\it \ }$^{3}${\it NAPP Group, Indian Institute of Astrophysics, }\\
{\it Koramangala, Bangalore - 560~034, India}\\
$^{4}${\it School of Engineering and Applied Science, }\\
{\it University of Virginia}, {\it Charlottesville, VA 22904-4746, USA}\\
$^{5}${\it Department of Physics, University of Virginia, }\\
{\it P.O.Box 400714, Charlottesville, VA 22904-4714, USA}}
\title{An origin of the Universe determined by quantum physics and relativistic
gravity }
\maketitle

\begin{abstract}
We discuss the evolution of the Universe from what might be called its
quantum origin. We apply the uncertainty principle to the origin of the
Universe with characteristic time scale equal to the Planck time to obtain
its initial temperature and density. We establish that the subsequent
evolution obeying the Einstein equation gives the present temperature of the
microwave background close to the observed value. The same origin allows the
possibility that the Universe started with exactly the critical density, $%
\Omega =1,$ and remained at the critical density during evolution. Many
other important features of the observed Universe, including homogeneity and
isotropy, Hubble's constant at origin, its minimum age, present density etc.
are all predictions of our theory. We discuss also the testability of our
theory.
\end{abstract}

The present standard model of cosmology relies on a big bang origin of the
Universe in which reverse evolution via the Einstein equations yields a
formal singularity. There are physical reasons to believe that the birth of
the Universe was controlled by quantum processes which might avoid such a
singularity, but a satisfactory theory of quantum gravity is still to be
found.

In this paper we avoid the necessity of a formal quantum gravity theory to
describe the origin of the Universe by making the assumption that the
quantum origin would obey the uncertainty principle and relativistic gravity
in an essential way. This allows us to deduce some important conclusions
regarding observable features of the present Universe.

Quantum processes such as state life times, tunneling etc. are characterized
by the energy-time uncertainty relation, $\Delta E\Delta t\geq \hbar /2.$ In
the present discussion we are concerned with a physical system in its lowest
energy state and the energy itself serves as $\Delta E.$

Our fundamental hypothesis is that the Universe took a quantum birth with a
characteristic time scale of the Planck time, the time scale being fixed by
quantum physics and relativistic gravity. (This time scale arises naturally
if the origin involves crossing above a gravitational barrier, classically
forbidden. Then the barrier height is given by a condition like $%
2G(E/c^{2})/c^{2}r\simeq 1,$ with $r=ct$ and a second equation is provided
by the quantization condition $Et\simeq \hbar /2$. These can be solved for $%
t=$ $(\hbar G/c^{5})^{1/2}\equiv t_{P}\simeq 5.4\times 10^{-44}$ s). Then we
get the initial average energy as 
\begin{equation}
E_{i}\simeq \hbar /2t_{P}
\end{equation}

At the end of the creation process the Universe is filled with radiation and
relativistic constituents with equivalent temperature given by the average
energy divided by the Boltzmann constant, 
\begin{equation}
T_{i}=E_{i}/k_{B}\simeq 7\times 10^{31}\ ^{\circ }K
\end{equation}

In effect, our assumptions have fixed the initial temperature of the
Universe to be approximately the Planck temperature. This value is a minimum
temperature consistent with the uncertainty principle$.$

We expect that in the beginning an energy content of $E_{i}$ will be present
in a volume scaled to the Planck length, if our theory is consistent. The
scale comes into picture through the causal horizon at Planck time, $%
r_{H}\simeq ct_{P}$. We do not make any assumption on the actual size of the
Universe at origin, since this is not a well defined concept in general,
though the observable Universe is finite at any epoch. Starting from the
quantum mechanical estimate, we get for the initial energy density 
\begin{equation}
\rho _{i}=\frac{1}{2}(\frac{\hbar c^{5}}{G})^{1/2}/\frac{4\pi }{3}(\frac{%
\hbar G}{c^{3}})^{3/2}=\frac{3}{8\pi }\frac{c^{7}}{\hbar G^{2}}
\end{equation}

This is in fact the critical density as we will show later. We assume that
for larger time scales General Relativity is valid, at least to a good
approximation, and evolve the temperature and density forward in time using
the Einstein's equations.

The equations governing the evolution of the scale factor and the density in
the Universe are 
\begin{eqnarray}
\stackrel{.}{R}^{2}+k &=&\frac{8\pi G}{3}\rho R^{2} \\
\frac{d}{dR}(\rho R^{3}) &=&-3pR^{2}
\end{eqnarray}

The equation of state $p=p(\rho )$ will allow the determination of the
evolution of the density as a function of the scale factor. With this, $R(t)$
can be determined from eq. 3. Here we consider a Universe that has been
radiation dominated from the beginning to until about the time $t\simeq
10^{12}$ s, and then matter dominated until present, $t_{0}\simeq 5\times
10^{17}$ s \cite{weinberg,kolb}. Since the initial density is estimated to
be the critical density we set $k=0.$

During the radiation dominated era the scale factor varies as

\begin{equation}
R(t)\sim t^{1/2}
\end{equation}

The temperature of the radiation and relativistic particles varies as the
inverse of the scale factor.

\begin{equation}
T(t)\sim 1/R(t)\sim t^{-1/2}
\end{equation}

Therefore, at the end of the radiation dominated era ($t_{r}\simeq 10^{12}$
s), we have for the temperature,

\begin{equation}
T(t_{r})=T_{i}/(t_{r}/t_{P})^{1/2}\simeq 5\times 10^{4}~^{\circ }K
\end{equation}

Near the end of this era the energy density of the relativistic particles
and radiation falls below that of the density of Hydrogen and Helium in the
Universe. The subsequent evolution of the temperature is slightly faster,
since the scale factor varies as $t^{2/3}$ during the matter dominated era,
whereas the temperature still varies as the inverse of the scale factor. So,
during $t\simeq 10^{12}$ s to the present, $t_{0}\simeq 5\times 10^{17}$ s,

\begin{equation}
T(t)\sim 1/R(t)\sim t^{-2/3}
\end{equation}
\begin{equation}
T(t_{0})=T(t_{r})/(t_{0}/t_{r})^{2/3}\simeq 1.7~^{\circ }K
\end{equation}

This prediction of the present temperature, starting from an initial
temperature determined by the quantum mechanical origin we have chosen and
the constants relevant to quantum physics and relativistic gravity, is
remarkably close to the observed present temperature. The temperature of the
microwave background will be a factor of about 1.4 higher from our estimate
due to the entropy production in $e^{+}-e^{-}$ annihilation\cite
{weinberg,kolb}, which yields a final value of $2.4^{\circ }$ K$.$ A number
of other factors could slightly alter our estimate, such as the lack of
precision in knowledge of the age of the Universe, and the accurate use of
the transition from the radiation to the matter dominated era. Even the
presence of a cosmological constant that dominates the energy density during
the relatively recent evolution of the Universe would not change our
estimate considerably.

While the temperature of the background radiation is an important touchstone
for any hypothesis about the birth and the evolution of the Universe, it is
desirable that the hypothesis is tested with other comparisons that are
significant. It turns out that the Universe can start off naturally with the
critical density if its origin is in the quantum process we are discussing.
This feature is very attractive since inflation is the only other known
mechanism that can endow the Universe with critical density naturally,
without drastic fine tuning.

We had estimated above that the energy density at origin is $\rho _{i}=\frac{%
3}{8\pi }\frac{c^{7}}{\hbar G^{2}}.$ \ This is the critical energy density
as seen directly from the Friedman equation, 
\begin{equation}
\left( \frac{\stackrel{.}{R}}{R}\right) ^{2}+\frac{k}{R^{2}}=\frac{8\pi
G\rho }{3}
\end{equation}

For a Universe at critical density (flat), $k=0.$ At the origin we have $%
\stackrel{.}{R}/R=c/ct_{P}=1/t_{P}.$ Then we have 
\begin{eqnarray}
\frac{1}{t_{P}^{2}} &=&\frac{8\pi G\rho _{c}}{3} \\
\rho _{c} &=&\frac{3}{8\pi }\frac{c^{5}}{\hbar G^{2}}
\end{eqnarray}

The critical energy density is $\rho _{c}c^{2},$ which is just what we have
calculated from the quantum origin scenario. The density evolves as $1/t^{2}$
during the radiation dominated and the matter dominated era and this gives
the present (critical) energy density. 
\begin{equation}
\rho _{0}=\rho _{i}/(\frac{t_{0}}{t_{P}})^{2}=\frac{3c^{2}}{8\pi Gt_{0}^{2}}
\end{equation}

It is important that this can be obtained only if we use the time scale $%
t_{P}$ in the uncertainty principle. If we use the uncertainty principle at
any other epoch, arbitrarily chosen, we do not get the density of the
Universe as the critical density, nor the correct present temperature. This
should be considered far from coincidence.

This theory solves a fundamental enigma about the observed Universe. It is
well known that if the present Universe has its density of the order of the
critical density then in the past the density should have been even closer
to the critical density. The ratio $\Omega =\rho /\rho _{c}$ moves closer
and closer to unity as $\left| \Omega -1\right| \sim (1+z)^{-2}$ in the past%
\cite{kolb}. At Planck epoch this ratio \ should have been unity accurate to
a part in $\approx 10^{60}$. This kind of fine tuning was beyond
explanation till present, except in an inflationary scenario for which the
present density is the critical density irrespective of the pre-inflationary
era.

We note that the quantity $2GM/c^{2}r_{H}$ at the quantum birth is unity,
where $M$ is the mass contained in a volume of radius equal to the horizon
size. \ $E_{i}t_{P}\simeq \hbar /2$ for the quantum birth condition gives $%
E_{i}\simeq M_{P}c^{2}/2.$ We get

\begin{equation}
2GM/c^{2}r_{H}=GM_{P}/c^{2}L_{P}=1
\end{equation}

It is interesting to note that the present observable Universe also
approximately obeys this relation. The horizon size changes linearly in $t,$
and the present horizon size is $r_{0}\simeq ct_{0}\simeq 10^{28}cm.$ The
scale factor itself increases at a slower rate, and the primordial
elementary volume of Planck length scale would have evolved into a size of
only about $10^{-2}$ cm at present.

We now show that the quantity $GM_{H}/c^{2}r_{H}$ is preserved at its
initial value. The density of the Universe varies as $\rho (t)\sim 1/t^{2}$
at all times and the horizon size increases as $r_{H}\sim t.$ The mass
contained in horizon size $r_{H}$ is 
\begin{equation}
M_{H}\sim \rho (t)r_{H}^{3}\sim t
\end{equation}

This gives 
\begin{equation}
GM_{H}/c^{2}r_{H}=\rm{constant}
\end{equation}

Another significant feature is that the quantum origin and the present
temperature of the microwave background fix a lower limit to the age of the
Universe. Since the initial temperature we estimated was the minimum
temperature, the time duration for the unique trajectory connecting the
initial temperature and the present temperature gives the minimum age of the
Universe, and it is about $15$ billion years. Our hypothesis also fixes
Hubble's constant at the origin independently as $H_{i}=\stackrel{.}{R}%
/R=c/ct_{P}=1/t_{P}\simeq 1.9\times 10^{43}$ $s^{-1}.$ Since $%
\stackrel{.}{R}/R\sim 1/t$ subsequently, for any power law evolution of $R$,
the present value of Hubble parameter would be 
\begin{equation}
H_{0}=\frac{1}{t_{P}}(\frac{t_{P}}{t_{0}})=1/t_{0}
\end{equation}
Combined with the age determined as earlier this gives the present Hubble
parameter as about $67$ km/s/Mpc. (Several recent measurements of the Hubble
constant have been reported. \ Gibson et al., \cite{gibson} for instance,
find $H_{0}=68\pm 2\pm 5$ km s$^{-1}$Mpc$^{-1}$, where the first uncertainty
value is the random component and the second one is the systematic
component. \ The general range into which the recent measurements are
converging \cite{freedman}\ is 60 to 75 km s$^{-1}$Mpc$^{-1}$.) \ The
deceleration parameter at origin is also determined since $q_{0}=\Omega $
during the initial epoch when the Universe was radiation dominated.

Now we bring out another important conclusion from our theory. Since the
whole of the presently observable Universe is causally linked to the
primordial Planck volumes in which the same fundamental constants determined
the initial conditions, this theory predicts homogeneity and isotropy of the
bare Universe, and all deviations from such a state have to occur through
structure formation. The presently observable Universe of size scale $%
10^{28} $ cm would have been of size scale $10^{-3}$ cm at Planck epoch (due
to dominantly $t^{1/2}$ evolution of size scale) \cite{kolb}. We see that
though the horizon size at the Planck time is much smaller than the total
size of the Universe, by a factor of about $10^{30},$ homogeneity and
isotropy over the whole Universe is assured at early times. Thus homogeneity
and isotropy is the signature of homogeneity and isotropy of the fundamental
constants.

If the hypothesis is fundamentally correct then the spatial variation and
anisotropy of $G,c$ and $\hbar $ are constrained very precisely. Any spatial
variation of these constants from the values that determined the critical
density at Planck time would lead to regions of space that are very far away
from critical density at present. Since the total density contrast at all
scales at the end of the radiation dominated epoch is limited to about $%
10^{-5},$ and since $\left| \Omega -1\right| \sim (1+z)^{-2}$ during earlier
times \cite{kolb}, the allowed spatial variation can be estimated to be
about $10^{-60}$ at origin over a length scale larger than Planck length.
The length scale of each primordial elementary volume at present is about $%
10^{-2}$ cm. The significance of this is that laboratory observations or
astrophysical observations that can confirm a spatial variation or
anisotropy of even such a tiny amount over their accessible scales will test
our theory by a large margin.

Since the present temperature of the microwave background is precisely
determined, the theory requires some level of temporal constancy of the
fundamental constants $G,c,$ and $h$. It turns out that a fractional
variation of about $10^{-11}/$year or smaller is consistent with our theory.

The validity of our hypothesis can have very interesting and useful
implications for the theoretical understanding of the evolution of the
Universe. In effect, it fixes two points in the history of the evolution of
the Universe, its birth and the present, and the possible evolution
trajectories consistent with various observations can be very limited. This
should be contrasted with Big Bang cosmology in which various physical
quantities at the origin are infinite or zero. The two precisely known
points in the thermal history we have outlined encompass all the interesting
aspects of particle physics during the evolution of the Universe, from
String physics to low energy physics.

\medskip 

We conclude that our theory and analysis provides significant evidence that
the initial conditions for the evolution of the Universe from a nonsingular
origin was fixed by the fundamental constants of quantum physics and
relativistic gravity and that the observed background radiation and the
present density close to the critical density are the relics of the quantum
birth of the Universe. 

\medskip \bigskip 

\noindent {\bf Acknowledgments}: We thank Ramanath Cowsik for discussions on
entropy production in the Universe and on other aspects of particle
thermodynamics. Laboratoire Kastler Brossel is Unit\'{e} de Recherche de
l'Ecole Normale Sup\'{e}rieure et de l'Universit\'{e} Pierre et Marie Curie,
associ\'{e}e au CNRS. The non-accelerator particle physics program at the
Indian Institute of Astrophysics is supported by the Department of Science
\& Technology, Government of India.

\medskip

\noindent E-mail addresses: 

$^{a}$unni@tifr.res.in, unni@lkb.ens.fr, $^{b}$gtg@virginia.edu, $^{c}$%
rcr8r@virginia.edu

\end{document}